# Analysis of the Dick Effect for AI-based Dynamic Gravimeter

Wen-Zhang Wang[1,2], Xi Chen[1,4,*], Jin-Ting Li[1,2], Dan-Fang Zhang[1,2], Wei-Hao Xu[1,2], Jia-Yi Wei[1,2], Jia-Qi Zhong[1,3,†], Biao Tang[1], Lin Zhou[1,3], Jin Wang[1,3,4,‡], Ming-Sheng Zhan[1,3,4,]

[1]*State Key Laboratory of Magnetic Resonance and Atomic and Molecular Physics, Innovation Academy for Precision Measurement Science and Technology, Chinese Academy of Sciences, Wuhan 430071, China*
[2]*School of Physical Sciences, University of Chinese Academy of Sciences, Beijing 100049, China*
[3]*Hefei National Laboratory, Hefei, 230088, China*
[4]*Wuhan Institute of Quantum Technology, Wuhan 430206, China*

Atom interferometer (AI)-based dynamic gravimeter enable high-precision absolute gravity measurements, crucial for applications in geophysics, navigation, resource exploration, and metrology. Understanding their underlying mechanisms and minimizing measurement noise are essential for enhancing performance. This work investigates the gravity measurement noise in AI-based systems induced by the dead time of the classical accelerometer. Using actual dynamic gravity measurement data, we demonstrate that a dead time of 0.12 s introduces significant gravity measurement noise of 8 mGal. To elucidate the mechanism of this noise, we derive a formula for this noise in frequency domain, identifying high-frequency aliasing as its source. Analysis of the derived expressions indicates that reducing the dead time duration and suppressing the high-frequency noise of the acceleration are effective strategies for mitigating this noise. This work provides significant insights for noise analysis and future scheme design of AI-based dynamic gravimeters.

## I. INTRODUCTION.

Dynamic gravity measurement is an important way for obtaining gravity fields with high-precision and high-spatial resolution, with significant applications in geophysics [1], navigation [2], resource exploration [3,4], and metrology [5,6]. Dynamic atomic gravimeter is a hybrid gravimeter which combines the measurement of an atom interferometer and a classical accelerometer, which can achieve absolute gravity measurement with high-precision. The method of dynamic atom gravity measurement originates from the vibration compensation scheme in the static atom gravity measurement [7], which is used to reduce the vibration noise of the environment. Then this method is extended to the dynamic environment [8-14]. In 2011, Geiger et al. demonstrated the dynamic atom acceleration measurement in a 0g plane [15,16]. In 2013, Bidel et al. realized atom gravity measurement in an elevator [17]. In 2018 and 2020, Bidel et al. demonstrated dynamic atomic absolute gravity measurements on ship and airplane, which showed superior gravity measurement precision compared to the traditional spring dynamic gravimeter [18-20]. In 2022, Guo et al. demonstrated dynamic atom gravimeter on moving vehicle [21,22]. In 2023 and 2025, Wu et al. demonstrated dynamic gravity measurement on ship and then in aircraft [23-27]. Our research group also developed a dynamic atom gravimeter, and carried out dynamic gravity measurement on a lake in 2021 and on the sea in 2024[28,29]. The developed gravimeter showed precision at the level of sub-mGal.

Improving the precision and clarifying the measurement mechanisms are hot topics for AI-based dynamic gravimeter. Xu et al. proposed a method to optimize the transfer function of the accelerometer [30]. Zhou et al. proposed a method to suppress the cross-coupling effect [29]. Cheiney et al. proposed a Kalman-filter algorithm to reduce the measurement noise [31], which has been applied by several research groups for the dynamic gravity data processing [32,33]. Huang et al. modeled the temperature of accelerometer to suppress its drift [34].

The acceleration measured by AI is used to calibrate the classical accelerometer in the hybrid gravity measurement. The acceleration acquisition of the classical accelerometer is usually triggered to be synchronized to the AI's time sequence to fulfil the requirement of hybrid gravity measurement. However, the triggering process for the data acquisitions usually needs a time interval for the triggering preparation and data processing, which will result in incomplete data acquisition for a classical accelerometer. This problem is the Dick effect in AI-based dynamic gravimeter.

Dynamic gravity measurement extracts weak gravity signals from strong acceleration background. The magnitude of gravity variation is in the order of mGal ($10^{-5}$m/s$^2$), while the magnitude of acceleration background is usually in the order of m/s$^2$. The 5 orders suppression of acceleration noise is achieved by low-pass filtering (LPF). Missing acceleration data might introduce additional gravity measurement noise. While the Dick effect has been studied in atom clock field [35-40], its specific impact on AI-based dynamic gravity measurement remains unexplored. This work addresses the following questions: 1) Does the missing acceleration leads the gravity measurement noise? 2) If so, what is the physical mechanism for this noise, and how can its amplitude be calculated? 3) How can this noise be suppressed? Addressing these questions will yield deeper mechanistic insights and guide improved AI-based dynamic gravimeter design.

*Contact author: chenxi@apm.ac.cn
†Contact author: jqzhong@apm.ac.cn
‡Contact author: wangjin@apm.ac.cn

This article is organized as follows. In section II, the principle and the data processing procedure of AI-based dynamic gravity measurement are introduced. By introducing the dead time of the acceleration, the gravity measurement noise is calculated in time domain by using an actual dynamic gravity measurement data. In section III, by analyzing the frequency spectrum of acceleration, the formula of the dead time induced gravity noise is derived, and the physical mechanism for this noise is explained. In section IV, the dead time induced gravity noise is calculated by using actual dynamic gravity measurement data, and the noise distribution over frequency is illustrated. In section V, conclusion and discussion are given.

## II. TIME DOMAIN ANALYSIS OF DEAD TIME INDUCED GRAVITY MEASUREMENT NOISE

### A. Principle of AI-based dynamic gravity measurement

An AI-based dynamic gravimeter combines an AI and a classical accelerometer. The AI measures the acceleration using the free-falling cold atom cloud. This measured acceleration is absolute, however, its data are discrete and its measurement range is limited. The classical accelerometer is solidly connected to the AI, measures acceleration synchronously. This measured acceleration is continuous and its measurement range is large, however, it has unknown offset and drift. Combining both enables continuous, absolute gravity acceleration measurements with a wide range [41].

The principle of the hybrid gravity measurement is illustrated in Fig. 1(a). For the AI, the cold atom cloud experiences the $\pi/2$-$\pi$-$\pi/2$ Raman laser pulses to compose an interference loop. The population $P$ is measured by fluorescence detection. The classical accelerometer measures the acceleration of the AI at the same time with an unknown offset. This acceleration is defined as $a_{cla}$, and the value of $a_{cla}$ in the time interval of interference time, associated with a so-called sensitivity function [42], is used to calculate the compensation phase $\varphi_{com}$. Then $\varphi_{com}$ and $P$ are set as the x and y-axis to restore the interference fringe. Sine fitting is used to fit this fringe, and the obtained phase is defined as the offset phase $\varphi_{off}$. By using $\varphi_{off}$, the offset of acceleration of the classical accelerometer $a_{off}$ is obtained. The absolute acceleration $a_{abs}$ felt by AI is derived as $a_{abs}=a_{cla}-a_{off}$.

For the dynamic gravity measurement, $a_{abs}$ comprises three components: the gravitational acceleration $g$, the vibration acceleration $a_{vib}$ induced by the moving carrier, whose frequency is usually larger than 0.1 Hz, and the motion acceleration $a_{mot}$ of the carrier, whose frequency is determined by the motion of the carrier and cannot be separated from the gravity change in frequency domain. For the vibration acceleration $a_{vib}$, by using low-pass filtering with a filtering time in the order of 100 s, this acceleration component is filtered out. For the motion acceleration $a_{mot}$, by using the position information of the carrier obtained by the Global Navigation Satellite System (GNSS), this acceleration component can be calculated. The gravitational acceleration g is obtained as $g=a_{abs}-a_{vib}-a_{mot}$.

This hybrid dynamic gravity measurement process requires a continuous classical acceleration sequence, typically obtained by averaging $a_{cla}(t)$ over AI measurement cycles. For the requirement of synchronization of the hybrid gravity measurement, the acquisition of the classical acceleration is usually triggered by the AI's time sequence. Due to the triggering preparation and data processing, the measurement time $T_{mea}$ of $a_{cla}(t)$ within one experiment cycle does not fully cover the AI cycle time $T_{cyc}$ of a single AI experiment, as illustrated in Fig. 1(b). The missing sampling time $\Delta T=T_{cyc}-T_{mea}$ is defined as the dead time. This article investigates the gravity measurement noise induced by $\Delta T$.

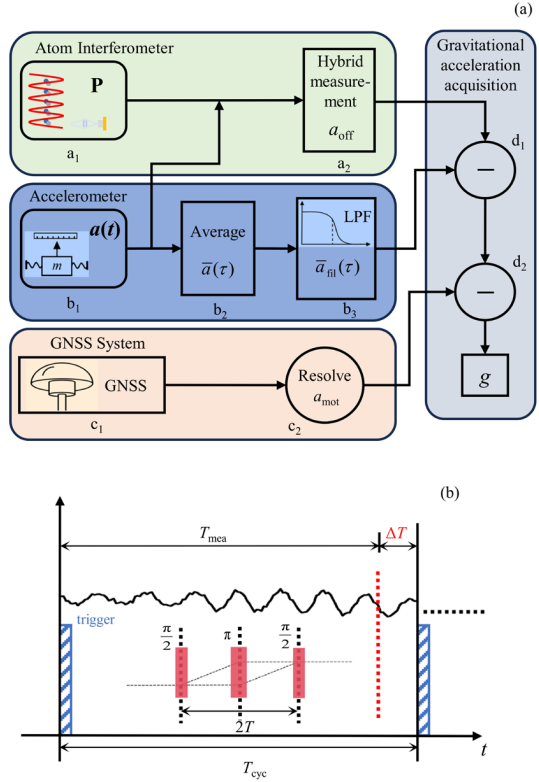

FIG 1. Principle of AI-based dynamic gravity measurement. (a) Gravity extraction procedure. (b) Acquisition of the classical acceleration $a_{cla}(t)$ with dead time $\Delta T$ within one AI experiment cycle.

### B. Calculation of gravity measurement noise

The AI-based dynamic gravimeter developed by our research group is shown in Fig. 2(a). An acceleration acquisition method is developed to avoid the dead time, which can collect all the acceleration data in an AI measurement cycle. Marine dynamic gravity measurements followed the procedure in Fig. 1(a), and the full-sampled acceleration is used to extract the gravity. The gravity survey is along a line with a speed of about 10 knots, and the length of the line is about 45 km. The measurement result can be found in Ref. [29], and a part of the measurement result is shown in Fig. 2. The acceleration $a_{cla}$ collected by the classical accelerometer is shown in Fig. 2(c), which has a peak to valley value of about 0.6 m/s². The measured gravity value is shown in Fig. 2(d). For comparison, a spring dynamic gravimeter is installed on the same ship (as shown in Fig. 2(b)), and measured the gravity synchronously. The measured gravity value is also shown in Fig. 2(d). The standard deviation of the gravity difference of these two gravimeters was 0.4 mGal. This gravity measurement result of the AI-based dynamic gravimeter in Fig. 2(d) has a high precision, and is used as a gravity reference $g_{ref}(t)$.

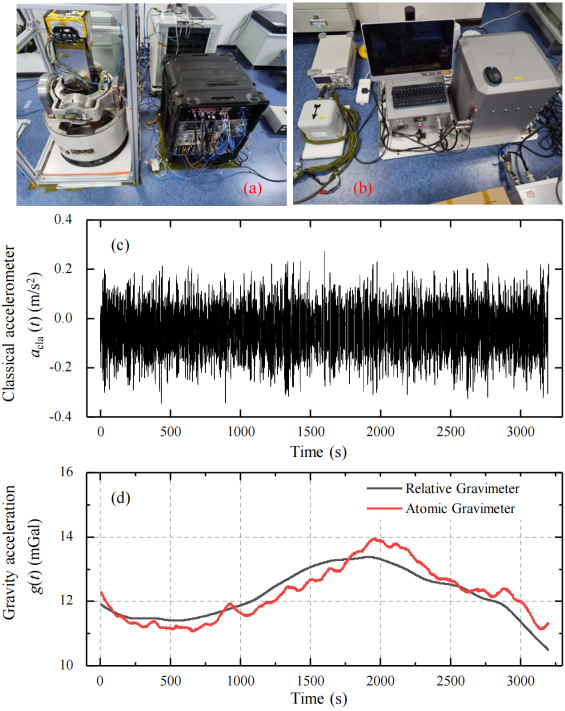

FIG 2. AI-based dynamic gravimeter and measurement data. (a) The AI-based dynamic gravimeter. (b) The spring dynamic gravimeter. (c) Acceleration $a_{cla}$ collected by the classical accelerometer. (d) The measured gravity value from both gravimeters. Both measured gravity values subtracted the same normal gravity model for the convenience to display.

For our AI-based gravity measurement experiment, the cycle time $T_{cyc}$ of AI is 0.6 s, and the sampling rate of the classical accelerometer is 250 Hz, yielding 150 acceleration data points in each cycle. To study the Dick effect in dynamic gravity measurement, acceleration data points with time $\Delta T$ are removed at the end of each cycle. Then, the procedure in Fig. 1(a) is followed to calculate the under-sampled gravity value $g_{und}(t)$. The gravity value $g_{und}(t)$ is compared with the $g_{ref}(t)$ to obtain the gravity difference $\Delta g(t)=g_{und}(t)-g_{ref}(t)$. The calculated $\Delta g(t)$ with different $\Delta T$ is shown in Fig. 3(a). The gravity difference fluctuated significant when $\Delta T$ increased. The standard deviation of $\Delta g(t)$ is noted as the gravity measurement noise $\sigma_{\Delta g}$, and the relationship between $\sigma_{\Delta g}$ and $\Delta T$ is shown in Fig. 3(b). When the dead time $\Delta T$ increased, the gravity difference fluctuation $\sigma_{\Delta g}$ increased too. $\sigma_{\Delta g}$ reaches 1 mGal even for a short dead time $\Delta T$=4 ms, and $\sigma_{\Delta g}$ increases to 8 mGal when $\Delta T$=100 ms. From this calculation, we find that the dead time causes non-negligible gravity measurement noise, and leads to severe degradation of gravity measurement precision.

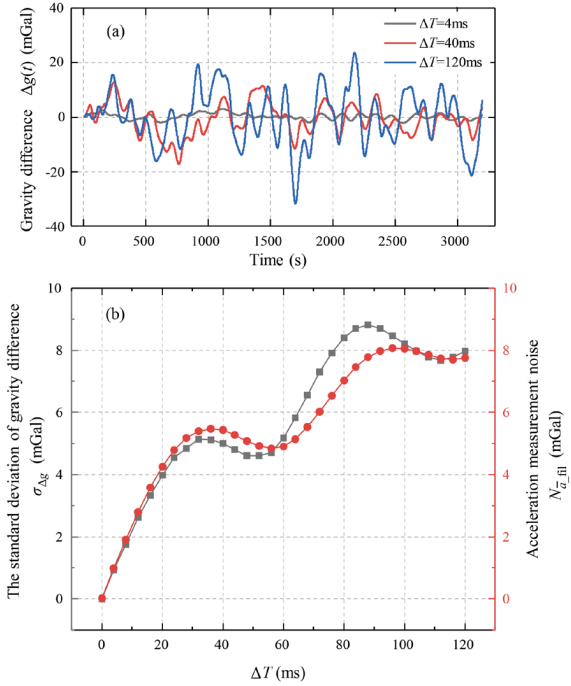

FIG 3. The gravity measurement noise caused by dead time $\Delta T$. (a) Gravity difference $\Delta g(t)$ between $g_{und}(t)$ and $g_{ref}(t)$. (b) Gravity measurement noise $\sigma_{\Delta g}$ over $\Delta T$ calculated in the time domain (black square), and increasing acceleration measurement noise $N_{\bar{a}\_fil}$ over $\Delta T$ calculated in the frequency domain (red circle).

## III. FREQUENCY DOMAIN ANALYSIS OF DEAD TIME INDUCED GRAVITY MEASUREMENT NOISE

### A. Principle of AI-based dynamic gravity measurement

Section II gives the time domain calculation of the gravity measurement noise $\sigma_{\Delta g}$, caused by dead time. However, the above calculation gives the magnitude of $\sigma_{\Delta g}$, but cannot provide a quantitative expression of $\sigma_{\Delta g}$, and cannot explain the mechanism of noise generation. Referring to Fig. 1(a), procedures b₁-b₃ are the only steps differing between cases with and without dead time. So, $\sigma_{\Delta g}$ is just induced by these three procedures. The acceleration signal in procedure b₃ is noted as $\bar{a}_{fil}(\tau)$, and the increased noise that add to $\bar{a}_{fil}(\tau)$ caused by the dead time is noted as $N_{\bar{a}\_fil}$, then $N_{\bar{a}\_fil}$ is just the same as $\sigma_{\Delta g}$. So, the mission is to derive the formula $N_{\bar{a}\_fil}$ from the classical acceleration $a_{cla}(t)$. For simplicity, we note $a_{cla}(t)$ as $a(t)$ in the following of the article.

Since $a(t)$ is random in time domain but has a stable form in the frequency domain. So, we start at the frequency domain signal of $a(t)$, which is noted as $A(v)$. The acceleration data are sampled through devices with a finite sampling rate. So, both $a(t)$ and $A(v)$ are discrete data sequence. We note them as $a(t_j)$ ($j=1…n$) and $A(v_i)$, ($i=1…n$), where $n$ is the number of acceleration data points, the time interval of $t_j$ is $T_{sam}$, which is sampling time of $a(t_j)$. The frequency interval of $v_i$ is $\Delta v$, which is equal to $1/nT_{sam}$, the range of $v_i$ is $(-1/2T_{sam}, 1/2T_{sam})$. The symmetry $|A(v_i)|=|A(-v_i)|$ allows subsequent calculations to consider only the non-negative frequencies $(0, 1/2T_{sam})$.

For the data used in this article, which is also shown in Fig. 2(c), we have $n=8\times10^5$, $T_{sam}=4$ ms, $\Delta v=3.125\times10^{-4}$ Hz. $a(t_j)$ can be extracted from $A(v_i)$ through inverse discrete Fourier transform (IDFT)

$$a(t_j) = \frac{1}{\sqrt{n}} \sum_{i=1}^{n} A(v_i) \cdot e^{-2\pi I(i-1)(j-1)/n}. \quad (1)$$

The measured acceleration $a(t_j)$ is then averaged in each cycle time $T_{cyc}$, the averaged acceleration is noted as $\bar{a}(\tau_k)$ ($k=1... m$), where $m=n/c$ is the number of the average acceleration data points, and $c=T_{cyc}/T_{sam}$ is the number of data points in each cycle. Because the existence of the dead time, $a(t_j)$ are only averaged in the measurement time $T_{mea}$, which has the form

$$\bar{a}(\tau_k) = \frac{1}{c_{mea}} \sum_{j=(k-1)c+1}^{(k-1)c+c_{mea}} a(t_j), \quad (2)$$

where $c_{mea}=T_{mea}/T_{sam}$ is the number of data points that averaged per cycle.

To eliminate the high frequency vibration noise, a low pass filter is applied on $\bar{a}(\tau_k)$. To calculate the signal after the low pass filter, which is noted as $\bar{a}_{fil}(\tau_k)$, we transfer $\bar{a}(\tau_k)$ to frequency domain, which is noted as $\bar{A}(f_l)$ ($l=1…m$), and set its value to be zero for frequency above the filter's cut-off frequency. $\bar{A}(f_l)$ has the form

$$\bar{A}(f_l) = \frac{1}{\sqrt{m}} \sum_{k=1}^{m} \bar{a}(\tau_k) \cdot e^{2\pi I(k-1)(l-1)/m}. \quad (3)$$

The interval of $f_l$ is $\Delta f=1/nT_{sam}$, and the range of $f_l$ is $(0, 1/2T_{cyc})$. Our objective is to calculate the increased noise $N_{\bar{a}\_fil}$ for the filtered signal $\bar{a}_{fil}(\tau_k)$, but not the signal $\bar{a}_{fil}(\tau_k)$ itself. For a random signal, the variance of the signal in time domain equals the sum of its power spectrum density (PSD) in frequency domain. So, we do not carry out the inverse Fourier transform of $\bar{A}(f_l)$ to obtain $\bar{a}_{fil}(\tau_k)$, but calculate the variance of $\bar{a}_{fil}(\tau_k)$ from $\bar{A}(f_l)$ directly. The PSD of $\bar{a}(\tau_k)$ has the form

$$S_{\bar{A}}(f_l) = \frac{1}{m} \cdot |\bar{A}(f_l)|^2. \qquad (4)$$

The variance of $\bar{a}_{fil}(\tau_k)$, which is noted as $\sigma_{\bar{a}\_fil}$, can be calculated as

$$\sigma_{\bar{a}\_fil}^2 = \sum_{l=0}^{l_{fil}} S_{\bar{A}}(f_l), \qquad (5)$$

where $l_{fil}=\lfloor nT_{sam}/T_{fil} \rfloor$ is the frequency index for the low pass filter's cut-off frequency $1/T_{fil}$, where the symbol $\lfloor\ \rfloor$ represents rounding downwards. For our parameter of the gravity data extraction, $T_{fil}=300$ s, and the corresponding $l_{fil}=10$. Eq. (5) gives the formal expression for the variance $\sigma_{\bar{a}\_fil}^2$ of the filtered acceleration $\bar{a}_{fil}(\tau_k)$. The procedure of derivation of this expression is shown in Fig. 4. We want to note that the standard derivation $\sigma_{\bar{a}\_fil}$ is not the noise $N_{\bar{a}\_fil}$ that we want to extract. This is because that beside the increased noise, $\sigma_{\bar{a}\_fil}$ also contain acceleration signal such as the gravity acceleration and motion acceleration.

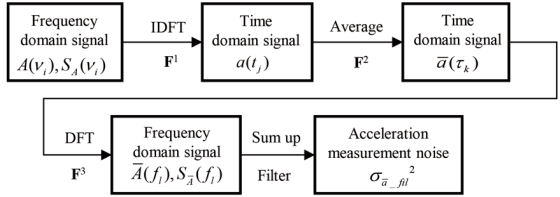

FIG 4. Procedures for deriving the PSD $S_{\bar{A}}(f_l)$ and the variance $\sigma_{\bar{a}\_fil}^2$ of the filtered acceleration $\bar{a}_{fil}(\tau_k)$.

## B. Simplification of the formal expression

Eq. (5) provides the formal expression of $\sigma_{\bar{a}\_fil}$ without separating noise from signal. In this chapter, we simplify Eq. (5) to give the analytic expression. The operation of IDFT, averaging, and DFT in Eq. (1)-(3) are represented as matrixes $\mathbf{F}^1$, $\mathbf{F}^2$ and $\mathbf{F}^3$, by substituting these matrixes into Eq. (3), we have

$$\bar{A}(f_l) = \sum_{i=1}^{n} D_{l,i} A(v_i), \qquad (6)$$

where, $\mathbf{D}=\mathbf{F}^3\mathbf{F}^2\mathbf{F}^1$ is defined as the transformation matrix. Substituting Eq. (6) into Eq. (4), we have

$$S_{\bar{A}}(f_l) = \frac{1}{m} \cdot \left|\sum_{i=1}^{n} D_{l,i} A(v_i)\right|^2 = c\sum_{i=1}^{n} H_{l,i} S_A(v_i), \qquad (7)$$

where $H_{l,i} = |D_{l,i}|^2$ and $S_A(v_i) = \frac{1}{n} \cdot |A(v_i)|^2$ is the PSD of the origin acceleration $a(t_j)$. Substituting Eq. (7) into Eq. (5), we have

$$\sigma_{\bar{a}\_fil}^2 = c\sum_{l=1}^{l_{fil}} \sum_{i=1}^{n} H_{l,i} S_A(v_i). \qquad (8)$$

Appendix A gives the derivation of the transformation matrix $\mathbf{D}$, which has the form

$$D_{l,i} = \frac{1}{n}\frac{\sqrt{c}}{c_{mea}} \frac{1-e^{\beta c_{mea}(i-1)}}{1-e^{\beta(i-1)}} \frac{1-e^{\beta n(i-l)}}{1-e^{\beta c(i-l)}}, \qquad (9)$$

where $\beta=-2\pi\mathrm{I}/n$.

## C. Properties of transformation matrix

The transformation matrix $\mathbf{D}$ has $m$ rows and $n$ columns. From Eq. (9) we can see that, the term $1-e^{\beta n(i-l)}$ in the numerator is constantly zero, and the term $1-e^{\beta c(i-l)}$ in the denominator is zero when $i=pm+l$, where $p=0,1\ldots,c-1$ is integer. We find that if $i=pm+l$, the limit for $D_{l,i}$ is not zero, and its value is derived in Appendix B. and for element $i \neq pm+l$, $D_{l,i}=0$. Then $H_{l,i} = |D_{l,i}|^2$ have the form

$$\begin{cases} H_{l,i} = \dfrac{1}{c}\dfrac{1}{c_{mea}^2} \dfrac{1-\cos[2\pi c_{mea}\frac{i-1}{n}]}{1-\cos[2\pi\frac{i-1}{n}]}; & i = pm+l, \\ H_{l,i} = 0; & i \neq pm+l. \end{cases} \qquad (10)$$

The matrix $\mathbf{H}$ can be expressed as a joint of $c$ diagonal matrixes $\tilde{\mathbf{H}}^p$, each matrix $\tilde{\mathbf{H}}^p$ has a size of $m\times m$. The matrix $\mathbf{H}$ and its sub matrix have the from

$$\begin{cases} \mathbf{H} = \{\tilde{\mathbf{H}}^0 \tilde{\mathbf{H}}^1 \cdots \tilde{\mathbf{H}}^p \cdots \tilde{\mathbf{H}}^c\} \\ \tilde{\mathbf{H}}^p_{l,l} = \mathbf{H}_{l,pm+l} \end{cases}. \qquad (11)$$

In Eq. (7), the PSD $S_A(v_i)$ has a length of $n$ and its frequency range is (0, $1/2T_{sam}$), $S_{\bar{A}}(f_l)$ has a length of $m$, and its frequency range is (0, 1/2$T_{cyc}$). So the role of sub-matrix $\tilde{\mathbf{H}}^p$ is to project the PSD element $S_A(\nu_i)$ in the frequency range ($p/T_{cyc}$, ($p+1$)/$T_{cyc}$) to the PSD element $S_{\bar{A}}(f_l)$ in the frequency range (0, 1/2$T_{cyc}$).

After the filtering, as shown in Eq. (8), the variation of $\sigma_{\bar{a}\_fil}^2$ only accumulates the element of $S_{\bar{A}}(f_l)$ for $l \leq l_{fil}$. So, the variation $\sigma_{\bar{a}\_fil}^2$ which $H_{l,i}(i=pm+l)$ contribute is for $pm<i<pm+l_{fil}$, which means only PSD element $S_A(\nu_i)$ in the frequency range ($p/T_{cyc}$, $p/T_{cyc}+l_{fil}/nT_{sam}$) contributes to $\sigma_{\bar{a}\_fil}^2$.

Now we illustrate the signal and increasing noise contribution in $\sigma_{\bar{a}\_fil}^2$, the signal of filtered acceleration $\bar{a}_{fil}(\tau)$ including the gravity acceleration and motion acceleration. These two accelerations are low frequency signals, The sub matrix $\tilde{\mathbf{H}}^0$ projects the PSD element $S_A(\nu_i)$ in low frequency range (0, $l_{fil}/nT_{sam}$), so, it contributes the signal. For $\tilde{\mathbf{H}}^p$ with $p>0$, their effect is to project high frequency signals of $S_A(\nu_i)$ to the $\sigma_{\bar{a}\_fil}^2$, so, they contribute the noise. Then the increasing noise $N_{\bar{a}\_fil}^2$ of the filtered acceleration $\bar{a}_{fil}(\tau)$ has the form

$$N_{\bar{a}\_fil}^2 = c \sum_{l=1}^{l_{fil}} \sum_{i=m+1}^{n} H_{l,i} S_A(\nu_i). \quad (12)$$

The value of $H_{l,i}(i=pm+l)$ for $p>0$ is shown in Fig. 5. For $\Delta T=0$, the value of $H_{l,i}$ for $pm<i<pm+l_{fil}$ is nearly zero. While for $\Delta T \neq 0$, $H_{l,i}$ for $pm<i<pm+l_{fil}$ deviates from zero, and it will cause noise according to Eq. (12). The mechanism for gravity noise $\sigma_{\Delta g}$, which is equal to the increasing noise $N_{\bar{a}\_fil}$, is the high frequency aliasing caused by dead time.

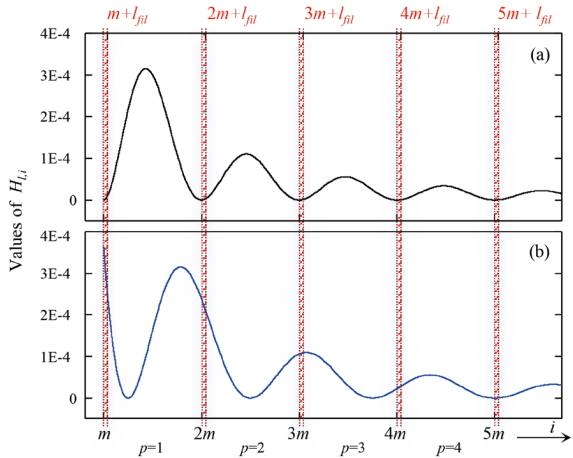

FIG 5. The value of $H_{l,i}(i=pm+l)$ for $p>1$. The red shaded area represents the range of $pm<i<pm+l_{fil}$. (a) $\Delta T=0$, (b) $\Delta T=0.12$ s.

## IV. ANALYSIS OF GRAVITY MEASUREMENT NOISE IN FREQUENCY DOMAIN

### A. Calculation and verification using actual data

To verify the correctness of the above derived equation, the actual acceleration data $a(t_j)$ in Fig. 2(c) is used to calculate its PSD $S_A(\nu_i)$, and the PSD $S_{\bar{A}}(f_l)$ of the averaged acceleration $\bar{a}(\tau_k)$ for $\Delta T=0.12$ s is calculated using Eq. (7), as shown in Fig. 6. The cut-off frequency of these two PSDs are 125 Hz and 0.83 Hz, respectively. These two PSDs exhibit similar shapes in frequency range (0.05 Hz to 0.83 Hz). This is because the dominated components of the PSD $S_A(\nu_i)$ is concentrated in this band, which is two order higher than its components in other frequency band. So, the aliasing effect caused by the dead time $\Delta T$ does not alter its shape. However, in the frequency range (DC-0.05 Hz), the PSD of $S_{\bar{A}}(f_l)$ is significantly higher and noisier than $S_A(\nu_i)$. This is just the aliasing effect cause by $\Delta T$. We check that, if $\Delta T=0$, $S_{\bar{A}}(f_l)$ and $S_A(\nu_i)$ are just the same in this low frequency band.

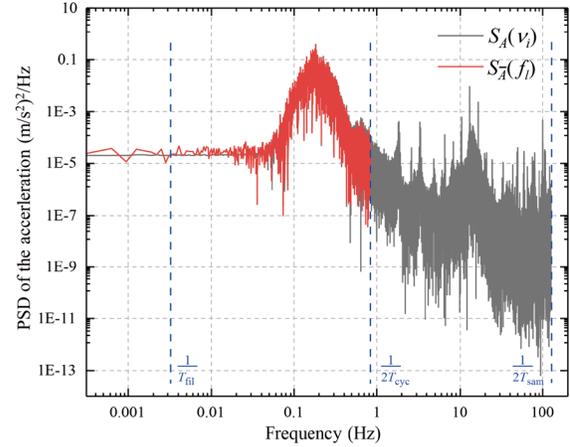

FIG 6. The calculated PSD $S_A(\nu_i)$ of the initial acceleration and the PSD $S_{\bar{A}}(f_l)$ of the averaged acceleration for $\Delta T=0.12$ s.

Then the increasing acceleration noise $N_{\bar{a}\_fil}$ is calculated using Eq. (12) with different $\Delta T$. The calculated result is shown in Fig. 3(b). As predicted in the beginning of Section III, $N_{\bar{a}\_fil}$ has the same amplitude as $\sigma_{\Delta g}$ that calculated in the time domain. The slight differences might be caused by the filtering method. For the time domain derivation for the σ$_{Δg}$, Bessel filter is used, while the frequency domain derivation of the $N_{\bar{a}\_fil}$, a square filter is used for convenience. However, the shape and amplitude of $N_{\bar{a}\_fil}$ and σ$_{Δg}$ coincide well, which verifies the correctness of the derivation equations in section III. From Fig. 3(b), we can see that if dead time is unavoidable, reducing its time duration is a way to alleviate its induced gravity measurement noise.

### B. Gravity noise distribution over frequency

The contribution of the sub-matrices $\tilde{\mathbf{H}}^p$ to the $N_{\bar{a}\_fil}$ is calculated to analyze the noise components in different frequency range. Noise components of $N_{\bar{a}\_fil}$ induced by $\tilde{\mathbf{H}}^p$ can be calculated as

$$N_{\bar{a}\_fil\_p}^2 = c \sum_{l=1}^{l_{fil}} \sum_{i=pm+1}^{(p+1)m} H_{l,i} \, S_A(\nu_i). \quad (13)$$

For $\tilde{\mathbf{H}}^p$, its corresponding frequency range is ($p/T_{cyc}$, $(p+1)/T_{cyc}$), which is about ($p\times 1.7$ Hz, $p\times 1.7+ l_{fil}\times \Delta\nu$ Hz) for our parameters. Fig. 7 shows the value of $N_{\bar{a}\_fil\_p}$ for $\Delta T$=0.12 s. The highest contribution is for $p$=2 and it has a value of about 7 mGal. Its corresponding frequency range is (3.3 Hz, 3.3+ $l_{fil}\times\Delta\nu$ Hz). Obvious peaks could be found at 1.7 Hz and 3.3 Hz in the PSD $S_A(\nu_i)$ in Fig. 6. So, reducing the high-frequency noise is another way to alleviate $N_{\bar{a}\_fil}$ if $\Delta T$ is unavoidable.

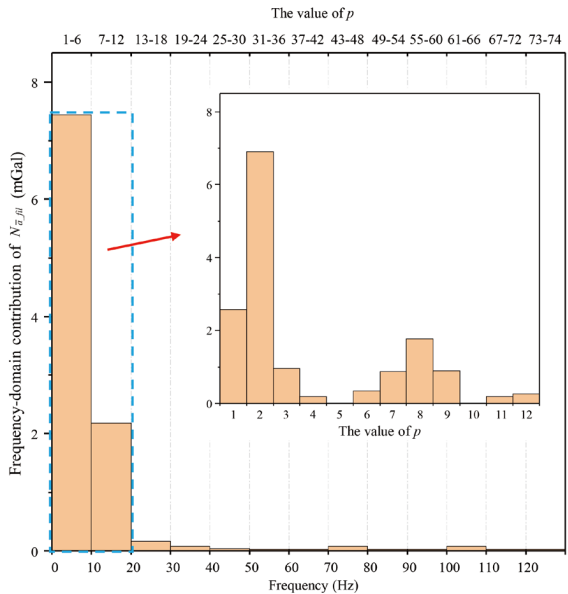

FIG 7. The calculated value of $N_{\bar{a}\_fil\_p}$ for $\Delta T$=0.12 s for different $p$. The inset figure is the enlarge picture for contributions in the frequency band (0~20 Hz).

### V. CONCLUSIONS AND DISCUSSION

This paper investigates studies on the AI-based dynamic gravity measurement noise induced by the Dick effect. By using the actual dynamic gravity measurement data from a marine gravity survey, we introduce the dead time for the classical acceleration and calculate its induced gravity measurement noise in time domain. To clarify the mechanism and find the formula for this noise, we start from the frequency domain signal of the classical acceleration, and derive the formula for the PSD of the average acceleration and increasing acceleration noise with dead time. Through analysis with the derived equations, we find that the high frequency aliasing is the mechanism that induces the gravity noise by dead time. Then actual classical acceleration data are used to calculate the PSD of the average acceleration and the increasing acceleration noise. The calculated noises in time domain and frequency domain coincide well, which verifies the correctness of the derived formulas. We find reducing the dead time duration and suppressing the high-frequency noise of the acceleration are two ways to reduce the gravity noise.

This work clarifies the mechanism, provides formulas and suppressing methods for the AI-based dynamic gravity measurement noise induced by the Dick effect, and it has significant importance both for the noise analysis and scheme design for the AI-based dynamic gravimeter that works in dynamic carriers such as ships, vehicles, and airplanes.

### ACKNOWLEDGMENTS

This work was supported by the Space Application System of China Manned Space Program (Second batch of the Scientific Experiment Project, JC2-0576), the Innovation Program for Quantum Science and Technology (2021ZD0300603, 2021ZD0300604), the Hubei Provincial Science and Technology Major Project (ZDZX2022000001), the Defense Industrial Technology Development Program (JCKY2022130C012), the National Natural Science Foundation of China (12174403,W2412045,12204493), the Natural Science Foundation of Hubei Province (2022CFA096), the Wuhan Dawn Plan Project (20230102010202082), The China Postdoctoral Science Foundation (2020M672453).

### AUTHOR DECLARATIONS
**Conflict of interest**

The authors have no conflicts to disclose.

## Author Contributions
**Wen-Zhang Wang:** Conceptualization (equal); Data Curation (equal); Formal Analysis (equal); Investigation (equal); Methodology (equal); Validation (equal); Writing - original draft (equal); **Xi Chen:** Conceptualization (equal); Formal Analysis (equal); Funding Acquisition (equal); Methodology (equal); Software (equal); Visualization (equal); Writing - review & editing (equal); **Jin-Ting Li:** Software (equal); Validation (equal); **Dan-Fang Zhang:** Data Curation (equal); Investigation (equal); **Wei-Hao Xu:** Formal Analysis (equal); Investigation (equal); **Jia-Yi Wei:** Investigation (equal); Validation (equal); **Jia-Qi Zhong:** Methodology (equal); Validation (equal); Visualization (equal); **Biao Tang:** Resources (equal); Visualization (equal); **Lin Zhou:** Resources (equal); Investigation (equal); **Jin Wang:** Project Administration (equal); Resources (equal); Supervision(equal); Writing - review & editing (equal); **Ming-Sheng Zhan:** Funding Acquisition (equal); Resources (equal); Supervision(equal); Writing - review & editing (equal);.

## DATA AVAILABILITY

The data that support the findings of this study are available from the corresponding author upon reasonable request.

## APPENDIX A

Directly calculation of the matrix $\mathbf{D}=\mathbf{F}^3\mathbf{F}^2\mathbf{F}^1$ is rather complicated. We proceed stepwise. First, calculate the matrix $\mathbf{F}^4=\mathbf{F}^2\mathbf{F}^1$, where the matrix element of $\mathbf{F}^1$ is $F^1_{j,i} = \frac{1}{\sqrt{n}} e^{\beta(i-1)(j-1)}$ and $\mathbf{F}^2$ represents the averaging process. Matrix multiplication yields:

$$\mathbf{F}^4 = \frac{1}{c_{mea}} \frac{1}{\sqrt{n}} \begin{pmatrix} F^4_{1,1} & \cdots & F^4_{1,n} \\ \vdots & F^4_{k,i} & \vdots \\ F^4_{m,1} & \cdots & F^4_{m,n} \end{pmatrix}, \quad (A1)$$

where the matrix element of $\mathbf{F}^4$ is $F^4_{k,i} = \frac{1}{c_{mea}} \frac{1}{\sqrt{n}} \frac{e^{\beta(k-1)(i-1)c}[1-e^{\beta(i-1)c_{mea}}]}{1-e^{\beta(i-1)}}$, then calculate $\mathbf{D}=\mathbf{F}^3\mathbf{F}^4$, where the matrix element of $\mathbf{F}^3$ is $F^3_{l,k} = \frac{1}{\sqrt{m}} e^{2\pi I(k-1)(l-1)/m}$, we have:

$$\mathbf{D} = \mathbf{F}^3\mathbf{F}^4 = \frac{1}{\sqrt{m}} \frac{1}{c_{mea}} \frac{1}{\sqrt{n}}$$
$$\begin{pmatrix} F^3_{1,1} & \cdots & F^3_{1,m} \\ \vdots & F^3_{l,k} & \vdots \\ F^3_{m,1} & \cdots & F^3_{m,m} \end{pmatrix} \begin{pmatrix} F^4_{1,1} & \cdots & F^4_{1,n} \\ \vdots & F^4_{k,i} & \vdots \\ F^4_{m,1} & \cdots & F^4_{m,n} \end{pmatrix}. \quad (A2)$$

Similarly, performing the matrix multiplication gives:

$$\mathbf{D} = \frac{1}{n} \frac{\sqrt{c}}{c_{mea}} \begin{pmatrix} D_{1,1} & \cdots & D_{1,n} \\ \vdots & D_{l,i} & \vdots \\ D_{m,1} & \cdots & D_{m,n} \end{pmatrix}, \quad (A3)$$

where the element of matrix $\mathbf{D}$ is:

$$D_{l,i} = \frac{1}{n} \frac{\sqrt{c}}{c_{mea}} \frac{1-e^{\beta c_{mea}(i-1)}}{1-e^{\beta(i-1)}} \frac{1-e^{\beta n(i-l)}}{1-e^{\beta c(i-l)}}; l \in [1,m], i \in [1,n] \quad (A4)$$

## APPENDIX B

The element of matrix $\mathbf{D}$ is shown in Eq. (A4), where the expression contains the factor $\frac{1-e^{\beta n(i-l)}}{1-e^{\beta c(i-l)}}$, which expands to $\frac{1-e^{-2\pi I(i-l)}}{1-e^{-2\pi I(i-l)c/n}}$. When $i=pm+l$, we have:

$$\frac{1-e^{-2\pi I(i-l)}}{1-e^{-2\pi I(i-l)c/n}} = \frac{1-e^{-2\pi I*pm}}{1-e^{-2\pi I*p}}. \quad (B1)$$

At this point, both the numerator and denominator of this factor are zero, now we add a tiny quantity $\delta$ to calculate its limit value. After expansion and calculation, we have:

$$\frac{1-e^{-2\pi I*p[m+\delta]}}{1-e^{-2\pi I*p[1+\delta]}} = \frac{1-e^{-2\pi I*pm}e^{-2\pi I*p\delta}}{1-e^{-2\pi I*p}e^{-2\pi I*p\delta}} = \frac{1-e^{-2\pi I*p\delta}}{1-e^{-2\pi I*p\delta}}. \quad (B2)$$

As a result, we have:

$$\lim_{\delta \to 0} \frac{1-e^{-2\pi I*p[m+\delta]}}{1-e^{-2\pi I*p[1+\delta]}} = 1. \quad (B3)$$

Therefore, when $i=pm+l$, $D_{l,i}$ has a non-zero value, whereas when $i \neq pm+l$, the denominator $1-e^{\beta c(i-l)}$ is non-zero while the numerator $1-e^{\beta n(i-l)}$ is zero, resulting in $D_{l,i}$ being zero.